# Bigravity : a bimetric model of the Universe.
# Exact nonlinear solutions.
# Positive and negative gravitational lensings.

**Jean-Pierre Petit and Gilles d'Agostini**

**jppetit1937@yahoo.fr**

___________________________________________________________________

**Abstract**

After a short summary of our bimetric model of the Universe, an exact nonlinear solution is built, which demonstrates the existence of solutions for our two coupled field equations system. In addition to the classical positive gravitational lensing, this solution is shown to also lead to *negative gravitational lensing*, a phenomenon previously described in 1995 (*Astrophysics and Space Science,* reference [10]). Such negative lensing provides an alternative interpretation for the observed faint magnitudes of high redshift galaxies, so far considered as dwarf galaxies.

___________________________________________________________________

## 1 – Introduction

In a previous paper [12] we evoked the principle of conjugated metrics, which we briefly summarize hereafter. Let us give the Universe a four-dimensional pseudo-riemannian manifold of signature (+ - - -). $T$ is a tensor density defined on this manifold, and $S^+$ is the Einstein tensor which is the gradient of the Hilbert functional in the metrics space, as:
(1)
$$H^+(g) = \int_M a R^+(g) \, dV_g$$

where $R^+$ is the scalar curvature. We can write Einstein's equation with a zero cosmological constant:
(2)
$$S^+ = \chi T$$

the solution of which is a metric named $g^+$ hereafter. Similarly, we call $g^-$ the conjugated metric solution of the equation:
(3)
$$S^- = -\chi T$$

and we thus have a system made of two coupled field equations.

## 2 – Exact solution as conjugated metrics.

Now we build a solution for this system, which we suppose stationary. We particularize the tensor density $T$ by making it describe an portion of space where the matter density $\rho^+$ is constant inside a sphere of radius $r_o$. Then a solution to equation (2) is what is known as the interior Schwarzschild metric:

(4)
$$ds^2 = \frac{(1 - \frac{r_o^2}{\hat{R}^2})^{3/2}}{\sqrt{1 - \frac{r^2}{\hat{R}^2}}} c^2 dt^2 - \frac{dr^2}{1 - \frac{r^2}{\hat{R}^2}}$$
$$- r^2 (d\theta^2 + \sin^2\theta \, d\varphi^2)$$
$$\text{for } r \leq r_o \qquad \hat{R}^2 = \frac{3 c^2}{8 \pi G \rho^+}$$

Positive mass particles and zero-mass positive-energy particles will follow the geodesics built from this metric, however only if they do not interact with positive mass matter contained in the sphere of radius $r_o$, which could be the case if the particles are neutrinos because they interact very weakly with matter.

If all calculations are taken from start in order to build a solution to equation (3), it appears the expression is the same, where $\rho^+$ is changed into $-\rho^+$. Hence, we write the interior metric related to the negative energy species:

(5)
$$ds^2 = \frac{(1 + \frac{r_o^2}{\hat{R}^2})^{3/2}}{\sqrt{1 + \frac{r^2}{\hat{R}^2}}} c^2 dt^2 - \frac{dr^2}{1 + \frac{r^2}{\hat{R}^2}}$$
$$- r^2 (d\theta^2 + \sin^2\theta \, d\varphi^2)$$
$$\text{for } r \leq r_o \qquad \hat{R}^2 = \frac{3 c^2}{8 \pi G \rho^+}$$

Imagine that the sphere of radius $r_o$, containing positive mass matter, is surrounded by an *extremely rarefied medium*. We do not use the term "vacuum" *because the perfect vacuum does not exist in astrophysics*. A perfect vacuum implies a region of space where none material particle could be found. Even if we managed to eliminate any material residue, photons would still remain, constituting a *black body* and bring their contribution to the energy-momentum tensor. Thus in a strict sense, the equation with zero second member
(6)
$$S = 0$$

from which the "exterior Schwarzschild solution" is classically derived, *does not have any physical sense. It is only an approximate expression.* Outside of this sphere of radius $r_o$ the solutions for the system:

(7) $$S^+ \cong 0$$

(8) $$S^- \cong 0$$

are:

(9) $$ds^2 = (1 - \frac{2GM}{c^2 r}) c^2 dt^2 - \frac{dr^2}{1 - \frac{2GM}{c^2 r}} - r^2 (d\theta^2 + \sin^2\theta d\varphi^2)$$

(10) $$ds^2 = (1 + \frac{2GM}{c^2 r}) c^2 dt^2 - \frac{dr^2}{1 + \frac{2GM}{c^2 r}} - r^2 (d\theta^2 + \sin^2\theta d\varphi^2)$$

where $M$ is the matter amount inside the sphere of radius $r_o$. Taking that fact into account, the geodesic system is composed by two distinct sets :

- Geodesics from the metric $g^+$, interior, then exterior, which are paths followed by positive mass and energy particles (and by zero-mass but positive energy photons). There is no discontinuity

- Geodesics from the metric $g^-$, interior, then exterior, which are paths followed by negative mass and energy particles (and by zero-mass but negative energy photons)

As in the classical theory ( internal and external metrics $g^+$ ) the continuity of the geodesics built from the second metric ( internal and external metrics $g^-$ ) is ensured everywhere.

The paths of matter particles and photons can be integrated and calculated in both cases. For positive mass particles we get:

(11) $$\varphi = \varphi_0 + \int_{u_0}^{u} \frac{du}{\sqrt{\frac{c^2 l^2 - 1}{h^2} + \frac{2GM}{h^2} u - u^2 + 2GMu^3}}$$

a classical formula in which u is the inverse of radial distance, φ the polar angle, l and h are parameters related to the initial kinetic energy and angular momentum, respectively. Keplerian trajectories are found through a Newtonian approximation. For the positive energy photons (so-called null-geodesics ), we get (classical formula too):

(12)
$$\varphi = \varphi_0 + \int_{u_0}^{u} \frac{du}{\sqrt{\frac{c^2 l^2}{h^2} - u^2 + 2GMu^3}}$$

which in particular gives the classical (positive) gravitational lensing :

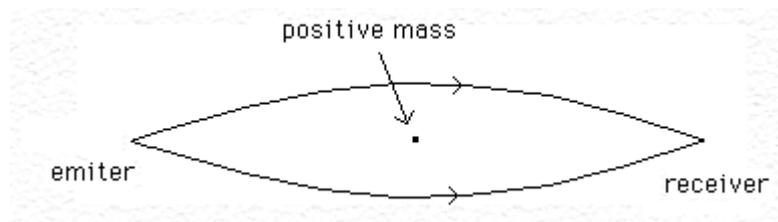

**Trajectories of positive energy photons, deviated by a positive mass M**

With the metric $g^-$ we can calculate the trajectories of negative mass particles and negative energy photons. Formulas become as below, first for negative mass particles the geodesics trajectories are also plane and given by :
(13)
$$\varphi = \varphi_0 + \int_{u_0}^{u} \frac{du}{\sqrt{\frac{c^2 l^2 - 1}{h^2} - \frac{2GM}{h^2} u - u^2 - 2GMu^3}}$$

and for the negative energy photons ( null-geodesics) :
(14)
$$\varphi = \varphi_0 + \int_{u_0}^{u} \frac{du}{\sqrt{\frac{c^2 l^2}{h^2} - u^2 - 2GMu^3}}$$

A positive mass M behaves like a repulsive object for particles having a negative mass and energy. It also creates a *negative* gravitational lensing, presented in 1995 in reference [9].

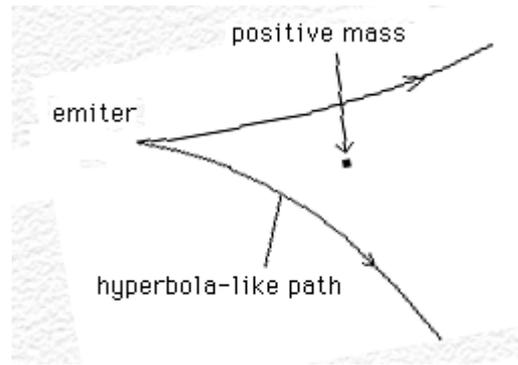

**Trajectories of negative energy photons, deviated by a positive mass M**

### 3 – Geodesics near a negative mass

The problem is completely symmetrical. Let us imagine an extremely rarefied space region, except inside a sphere of radius $r_o$ filled with negative mass matter.

As shown in the previous paper [12], this matter being self-attractive can form conglomerates of mass *M* and we can suppose they are filled with negative mass matter with constant density $\rho^-$. As a convention we decided to represent the *absolute value* of this negative mass density. Thus for the interior metric $g^+$ we get:

(14)
$$ds^2 = \frac{(1 + \frac{r_o^2}{\hat{R}^2})^{3/2}}{\sqrt{1 + \frac{r^2}{\hat{R}^2}}} c^2 dt^2 - \frac{dr^2}{1 + \frac{r^2}{\hat{R}^2}} - r^2(d\theta^2 + \sin^2\theta \, d\varphi^2)$$

$$\text{for } r \le r_o \qquad \hat{R}^2 = \frac{3 c^2}{8 \pi G \rho^+}$$

and for the interior metric $g^-$ :

(15)
$$ds^2 = \frac{(1 - \frac{r_o^2}{\hat{R}^2})^{3/2}}{\sqrt{1 - \frac{r^2}{\hat{R}^2}}} c^2 dt^2 - \frac{dr^2}{1 - \frac{r^2}{\hat{R}^2}} - r^2(d\theta^2 + \sin^2\theta \, d\varphi^2)$$

$$\text{for } r \le r_o \qquad \hat{R}^2 = \frac{3 c^2}{8 \pi G \rho^+}$$

While for the exterior metrics:
- corresponding to positive energy species:

(16)
$$ds^2 = (1 + \frac{2GM}{c^2 r}) c^2 dt^2 - \frac{dr^2}{1 + \frac{2GM}{c^2 r}} - r^2 (d\theta^2 + \sin^2\theta d\varphi^2)$$

- corresponding to negative energy species:

(18)
$$ds^2 = (1 - \frac{2GM}{c^2 r}) c^2 dt^2 - \frac{dr^2}{1 - \frac{2GM}{c^2 r}} - r^2 (d\theta^2 + \sin^2\theta d\varphi^2)$$

What is interesting here is that a negative mass and energy concentration somewhere in space must create a negative gravitational lensing on background objects:

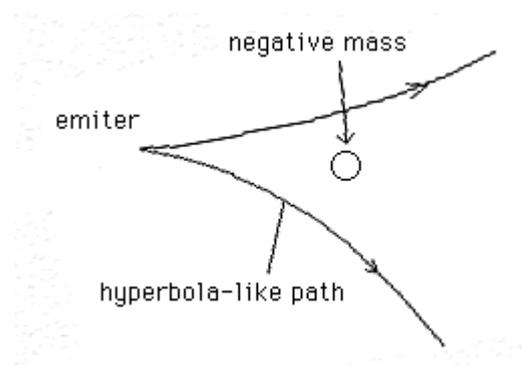

**Trajectories of positive energy photons, deviated by a negative mass**

**Nota bene**:

As shown in the preceding paper matter with negative mass ( and negative energy ) is self-attractive through gravitational force. Two negative masses attract each other through Newton's law. Opposite masses repel each other through (anti-Newton's law).

But positive mass and negative mass cannot interact through electromagnetic force. That's therefore we cannot detect negative mass through optical observation. For an observer made of positive energy matter the negative energy matter is a new kind of " black matter ". Its existence can be evidenced only through gravitational effects, gravitational lensing. To sum up :

- The positive energy mass concentrations produce positive lensing effect on positive energy photons, emitted by positive energy masses, measured by positive energy receivers and observers ( classical gravitational lensing effect ).

- The negative energy mass concentrations produce *positive* energy lensing effect on negative energy photons, emitted by negative masses, measured by negative energy receivers and observers ( if they exist ).

- The negative energy mass concentrations cannot be optically observed by positive mass receivers or positive mass observers, because such receivers or observers cannot receive negative energy photons, emitted by such negative mass concentrations.

- The positive energy mass concentrations cannot be optically observed by negative mass receivers or negative mass observers ( if they exist ), because such receivers or observers cannot receive positive energy photons, emitted by such positive mass concentrations.

- The positive energy photons, emitted by positive masses experience a *negative* gravitational lensing, due to the presence of negative mass concentrations.

- The negative energy photons, emitted by negative masses experience a *negative* gravitational lensing, due to the presence of positive mass concentrations.

For receivers or observers made of positive matter, negative masses behave like invisible objects.

For receivers or observers ( if they exist ) made of negative matter, positive masses behave like invisible objects.

*We can assume that negative masses could interact through weak and strong forces and form negative energy nuclei, atoms, molecules. Following this idea The world of negative mass and negative energy particles would be completely symmetric from ours in the way these particles interact. But its structures would be fully different. It would have no stars, no heavy atoms, no living beings, as it will be discussed in another paper.*

*The interaction between both worlds would only exist through gravity.* Notice the incident object (mass particle or photon) could cross the deflecting object which has opposite mass and energy, and we can draw:

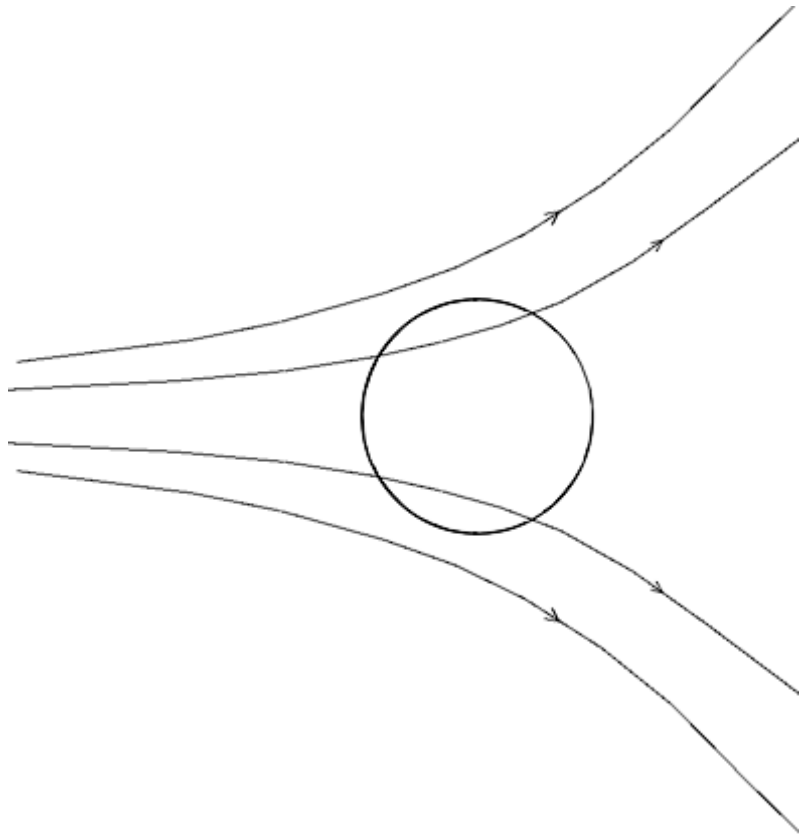

**Modification of photon trajectories
by a conglomerate made of opposite energy particles.**

This would reduce the apparent flux and size of background objects, exactly like a diverging lens does.

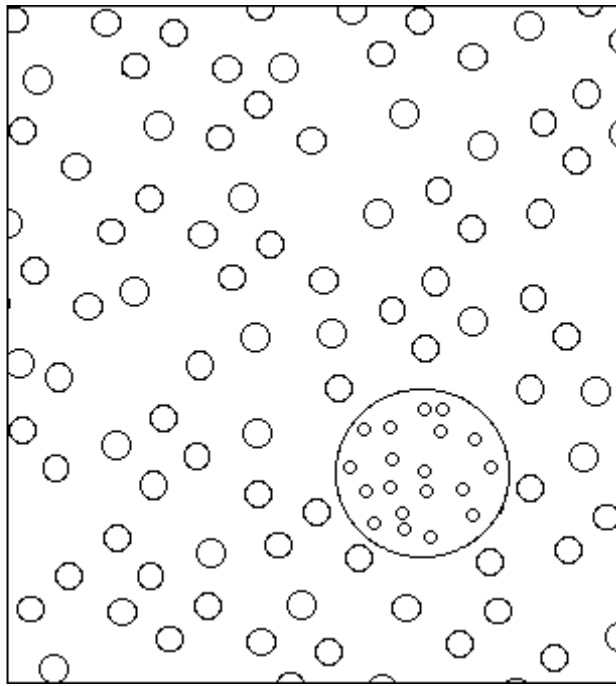

**Modification of background image of high redshift objects through negative
gravitational lensing created by a foreground negative mass conglomerate.**

Such phenomena *are effectively observed:* the farthest galaxies are seen as "dwarf", i.e. with fainter magnitudes because of invisible negative mass conglomerates along the light ray paths.

**4 – On the difficult mental picturing of a bimetric Universe**

We could decide to avoid any mental picturing and accept to sense the Universe only through metrics and equations. But there are different ways to imagine a world populated by two kinds of matter which would not optically see each other. For example, the draughts. The game is played only on black squares, so half the draughtboard is unemployed. One can imagine that two other players would play another game concurrently on the same board, but using the white squares. Imagine a draughts club having a room for "Men's Singles" and another for "Women's Singles". But due to lack of place the club direction decides the championships will now take place in the same room and on unique draughtboards, the men playing on the black squares and women on the white ones. To prevent any trouble, they are even fitted with polarized glasses, and polarized lights illuminate the pieces in order for the players to see only theirs, being unaware of what is played in the perpendicular game.

Pushing this idea to the limits we could imagine a "3D draughts" with black and white cubes :

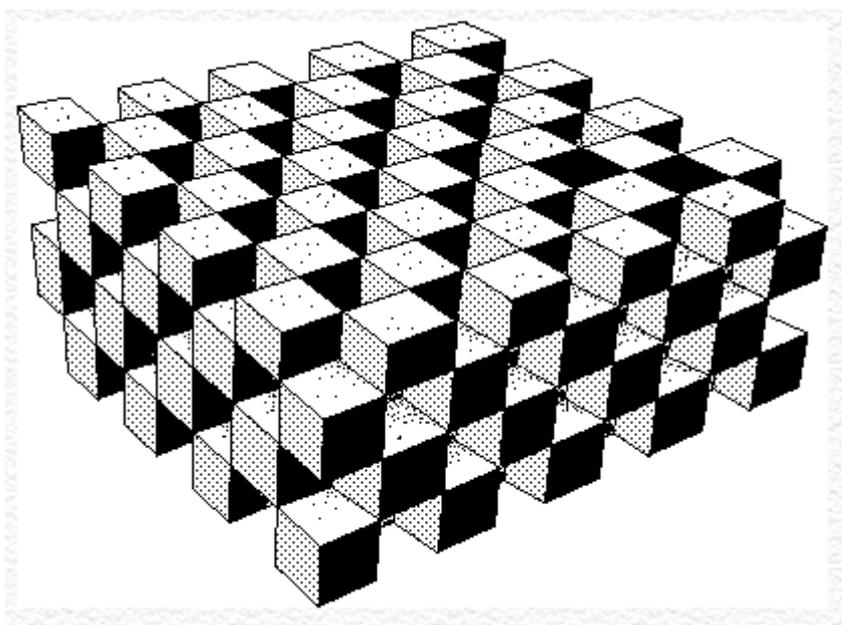

**3D draughts**

In a book published in 1997 in French, entitled "*On a perdu la moitié de l'univers*" (*We have lost one half of the Universe*) I even suggested a mental image in which the game would be played on an elastic flabby draughtboard, so as players would perceive the presence of other pieces, but without seeing them.

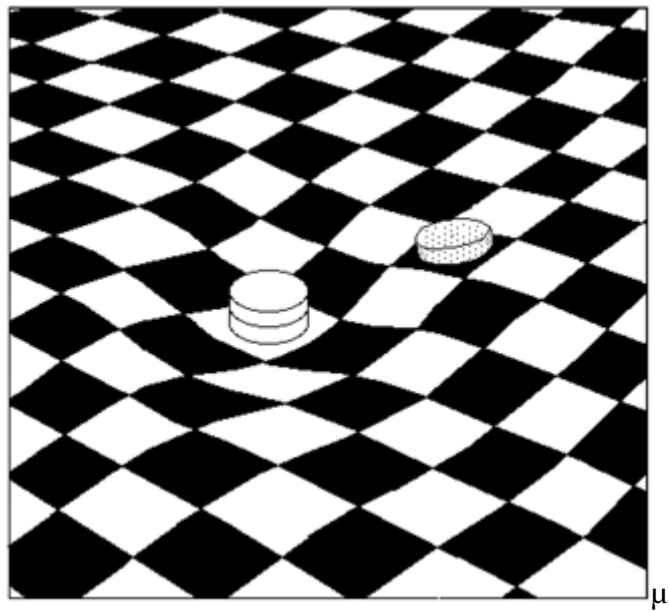

**The "flabby draughtboard"**

This would be a way to suggest the contribution of some (positive) "dark matter" to the gravitational field. But it does not solve the problem of *two metrics*. As we saw it, two points A and B, stationary in space (i.e. comoving) give *two distinct length measurements AB depending on the metric used*.

It is impossible to give a mental image with a single space framework. Consider a 2D word. The solution consists to show three surfaces. The first is a plane, it is the Euclidean Universe which is our intuitive mental picturing. We are completely inside a Platonic vision, in the well known allegory of the cave. We are constrained to adopt this point of view about things. Indeed, for any cosmic configuration there are always two universe patterns:

- *The perception of an observer that would be made of positive matter*
- *The perception of an observer that would be made of negative matter*

Perceptions, measurements of lengths are different. We are like inhabitants of Plato's cave, observing "Euclidean shadows" of phenomena inscribed into two structured hypersurfaces, one with the metric $g^+$ and the other with the metric $g^-$.

So let us consider a universe whose "bigeometry" is conditioned by contents of

- Energy-matter with positive value, related to tensor $T^+$
- Energy-matter with negative value, related to tensor $T^-$

according to the coupled field equations system ( [9], [10], [11], [12] ) :
(19)

$$S^+ = \chi \, ( T^+ - T^- )$$

$$S^- = \chi \, ( T^- - T^+ )$$

the scalar curvatures $R^+$ and $R^-$ are *opposite*. This is the very basis of our *conjugated geometries*, or *conjugated metrics*, a characteristic that allows us to suggest a didactic model providing a good mental picturing.

Let us give a *two dimensional image* of conjugated hypersurface.

Everybody knows how to build a cone, by cutting an angle $\theta$ into a disc, then resticking both edges. We shall see farther why we call this object a *posicone*.

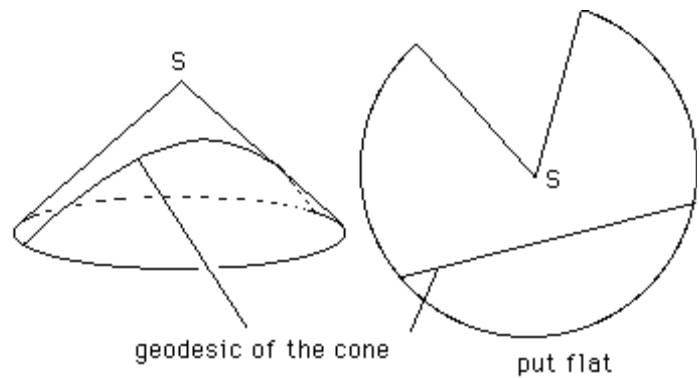

**Posicone**

If we draw a triangle thanks to geodesics, either the vertex *S* will not be outside that triangle (then the sum of its angles will be equal to $\pi$), or the vertex will be inside the triangle (and that sum will be equal to $\pi + \theta$); the Euclidean sum will be increased with the cut angle which enabled to create the cone. The vertex *S* is called a *concentrated curvature point*. But we can smooth the cone, *blunt* it, i.e. spread the curvature along a spherical cap. A boilermaker who wishes to build a blunt cone with two basic objects:

- A sheet cone with a concentrated curvature $\theta$
- A sphere

could manage to manufacture it without breaking geodesics, i.e. with no break in the tangent plane. He must know that the total curvature of a sphere $S_2$ is $4\pi$. So he starts to saw a spherical cap in order for it to contain an equivalent angular curvature. The cap surface is:
(20)
$$4\pi R^2 \frac{\theta}{4\pi} \quad i.e: \quad R^2 \theta$$

The angle $\theta$ is defined in radians. This makes possible to finalize the object. Then the artisan knows the cap perimeter, surface to edge, and puts it into his posicone to finalize the blunt posicone.

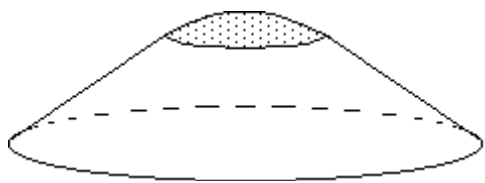

**Blunt posicone**

We have built a surface where the conical frustum has zero curvature and the spherical cap has a constant positive curvature. The following picture illustrates a cone going from a blunt state to a true conical shape, with a concentrated curvature point:

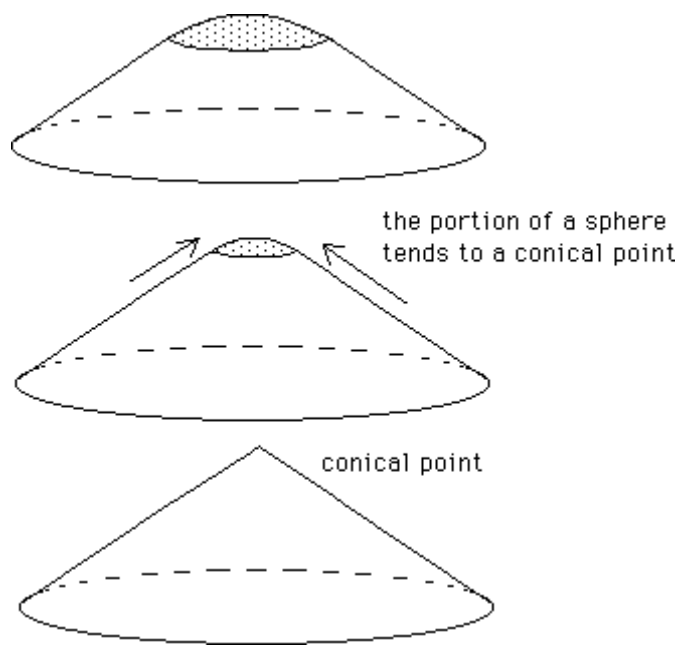

**Concept of a concentrated curvature point**

Let us now introduce our representation space, our Euclidean mental Universe, the "cave" upon which we represent this first family of objects identified by their trajectories:

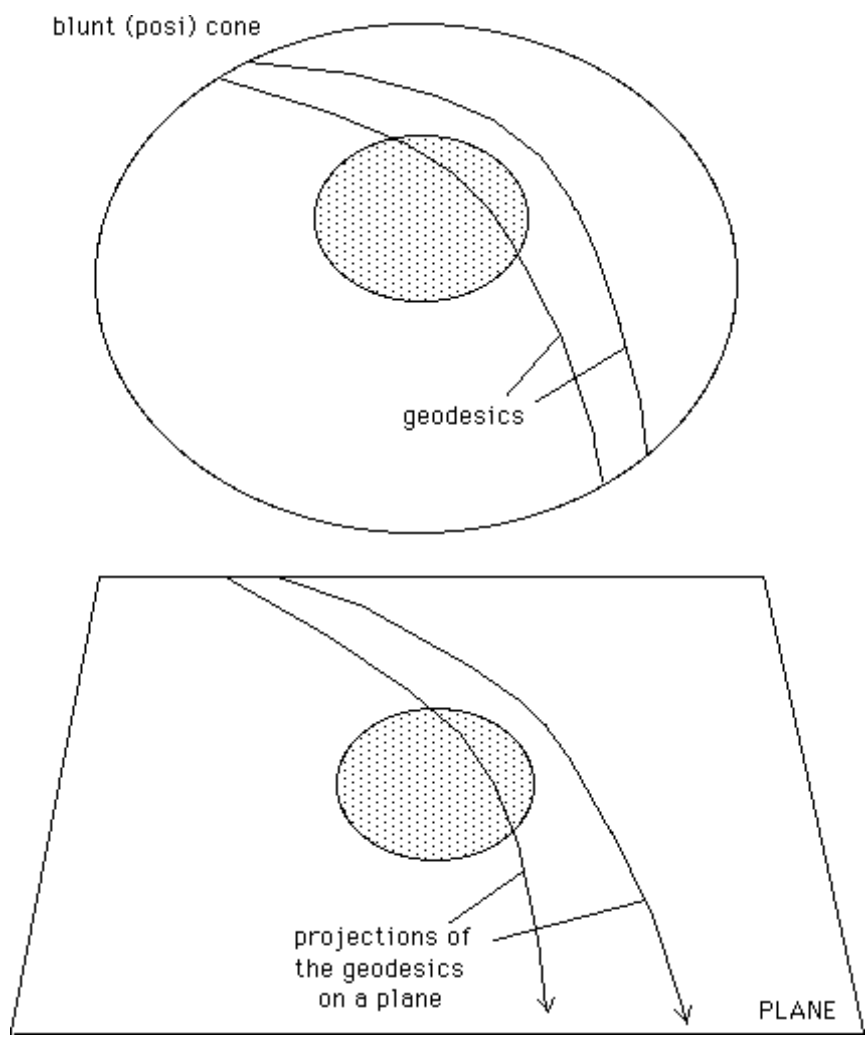

**Didactic image of the deflection of trajectories for positive mass objects
by a grayed out object also made of positive mass, constant density.
The trajectories crossing this mass would apply to neutrinos**

Then, how can we build a conjugated surface, having, point to point, an opposite curvature?
Let us start by making a *negacone*, whose vertex is a *negative curvature concentrated point*.

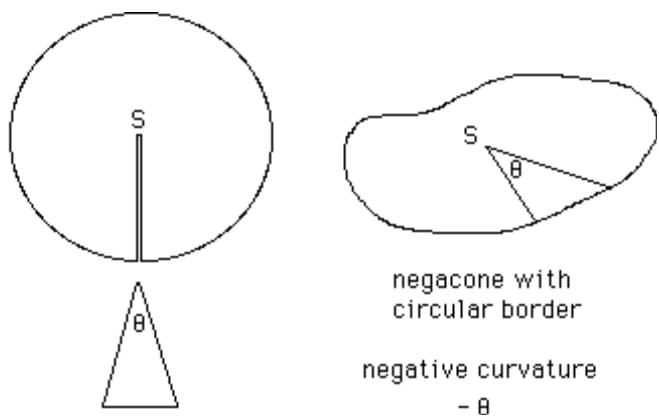

**Negacone : we *insert* a new angle sector $\theta$**

With the posicone it was easy to obtain a conical frustum by cutting the cone along a line made of points equidistant from the vertex. We can do the same thing with a negacone. Hereafter a negaconical frustum put flat on plane.

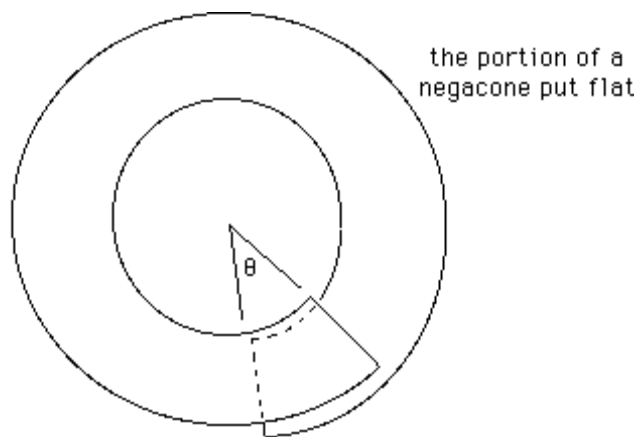

**Negaconical frustum, put flat**

The negacone is a relatively perplexing object, even when we have one in hands. It is never useless to build one and to handle it, even for an experienced geometer.
The *blunt negacone* will be built by connecting a horse saddle ( constant negative curvature surface ) on the negaconical frustum.

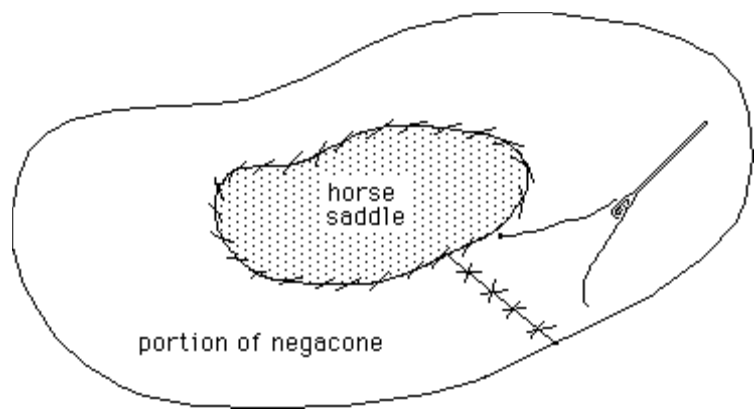

**Stitching between a negaconical frustum and a horse saddle**

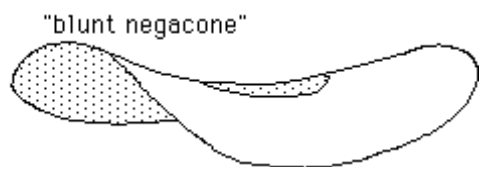

**The blunt negacone**

In the food industry, some little elements with a negative curvature surface do exist, these potato "negachips" can be purchased in supermarkets.

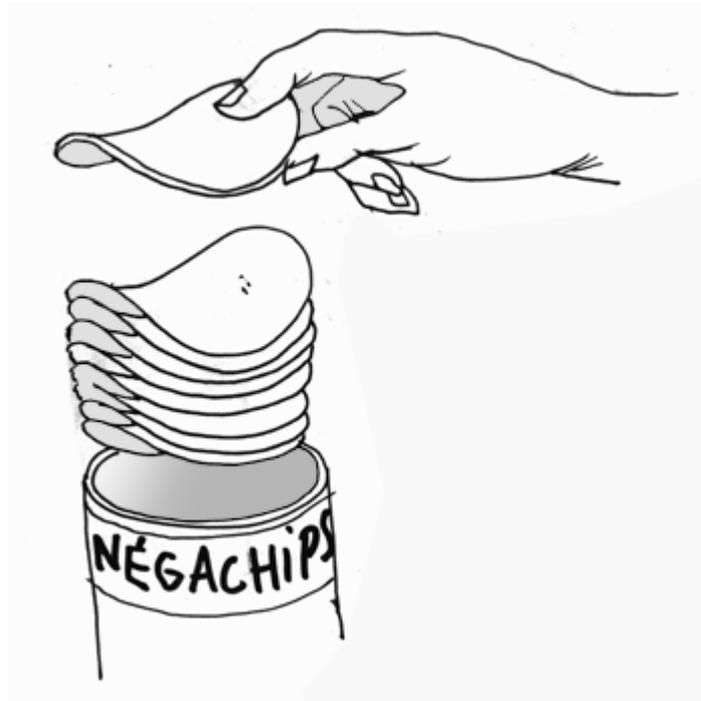

"**Negachips**"

How to measure the negative curvature quantity inside such a negachip? By sticking on its circular edge a negaconical frustum, trying to preserve the continuity of the tangent plane.

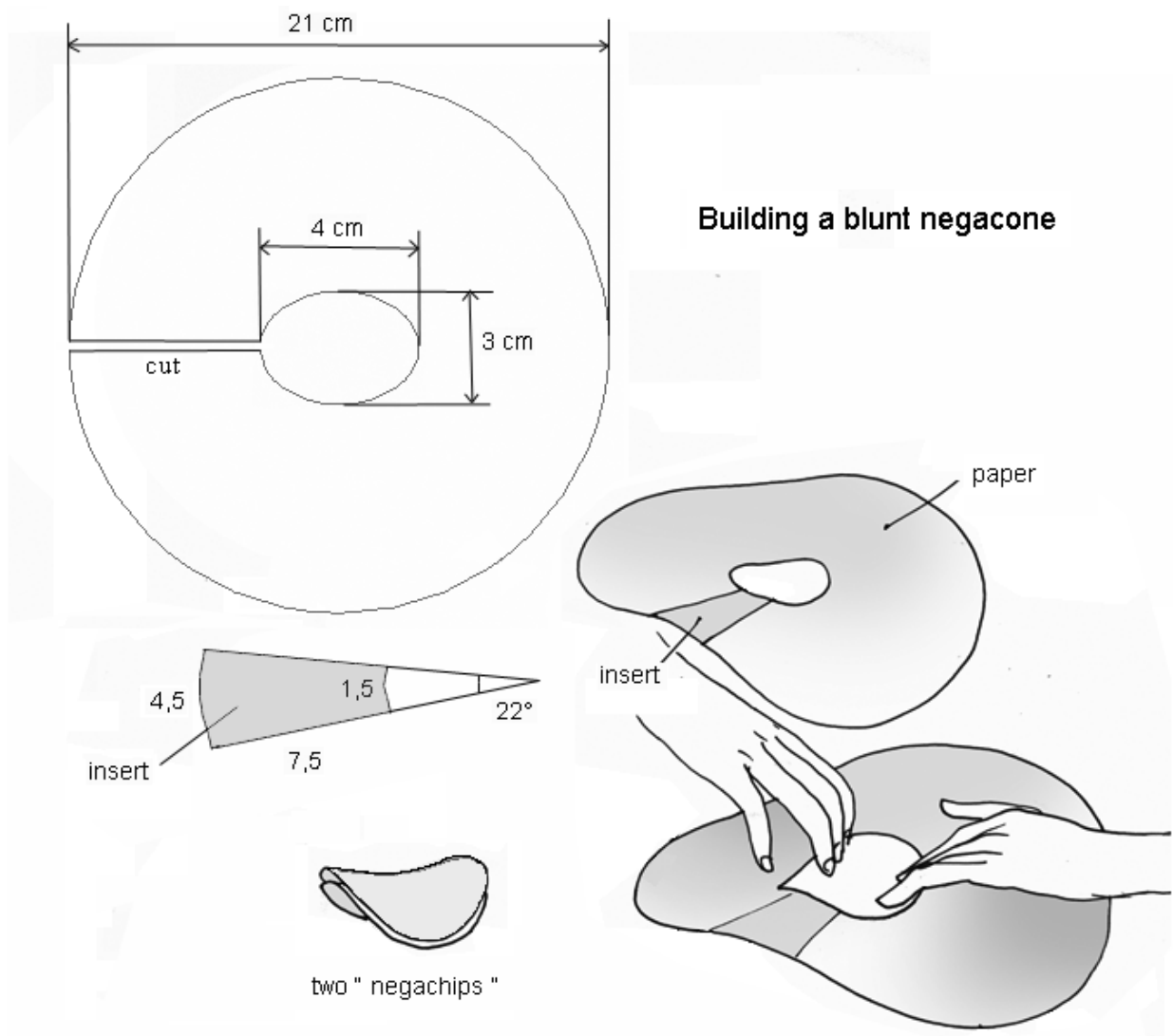

Finding some glue able to fix such a fatty object is difficult. Moreover the edge of these negachips is not circular. We can roughly determine the center of mass of the object then trace geodesic arcs up to its edge. These arcs will be as meridian arcs starting from the "pole" of the spherical cap, the experiment showing they have various lengths. This is why the central hole in the above illustration is elliptic and not circular, so as the negacone can be adjusted into the negaconical frustum.

This handling provides a rough estimate of the european negachip negative curvature:

$$\sim 22°$$

One could wonder if this experiment, a bit like a comic strip, is adequate in a scientific publication. Experience has shown not everybody sees well into space, and do not immediately understand 3D drawings. In this case the experiment helps a lot.

The opposite curvature shape would be a blunt posicone whose spherical cap contains a +22° curvature. Then we would have two *conjugated surfaces* having *opposite local curvatures*. They could be associated like in the illustration below.

*Thanks to that object, we are able to understand two conjugated surfaces.*

The first has an area with a constant positive curvature, surrounded by an Euclidean surface with no curvature. Then the conjugated surface is put in front of it, where the curvature is, point to point, opposite.

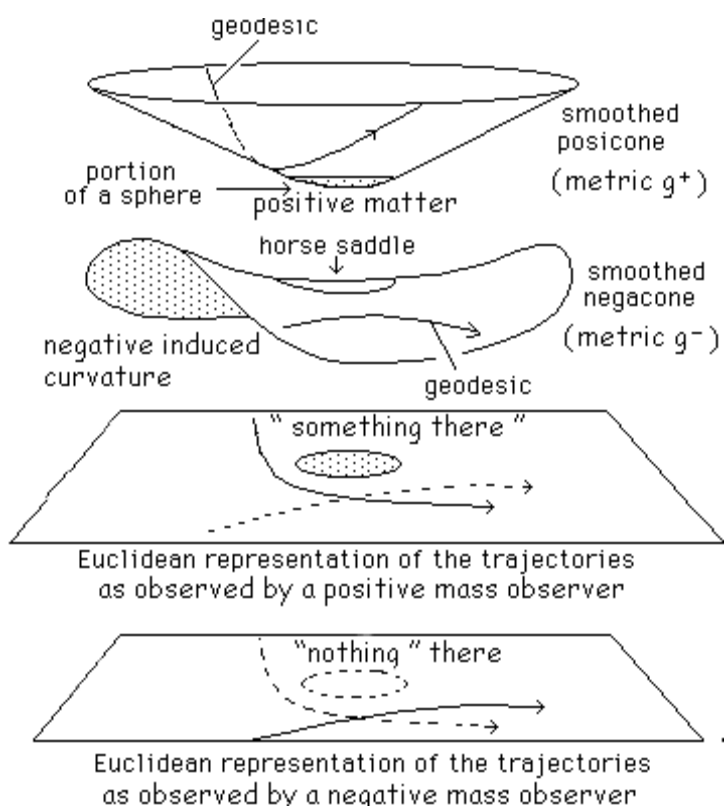

**Conjugated surfaces** (with opposite curvatures)

Suppose the upper surface is related to the metric $g^+$. For us, "Euclidean observers", matter trajectories we can see tell us the grayed area contains some attractive, positive mass concentration. From that metric we can calculate the geometric tensor $S^+$ and deduce the metric $g^-$ from the relation
(21)
$$S^- = -S^+$$

From the curvature field of the first surface, we can deduce the *conjugated curvature field* of the second one.

Then we can imagine the opposite situation: the matter is supposed to be now located in the second surface, upon which are inscribed the trajectories of negative energy particles.

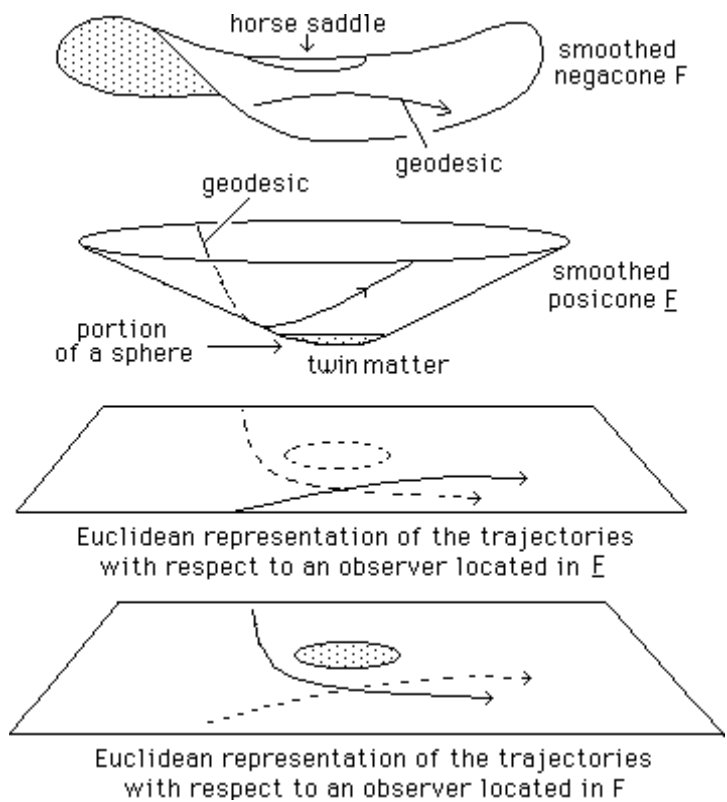

**Bottom: observers' perceptions in their "Euclidean mental space", one who is made of positive mass and the other of negative mass.**

The following drawing illustrates the negative lensing caused by some (invisible) negative mass conglomerate (located inside the dotted circle).

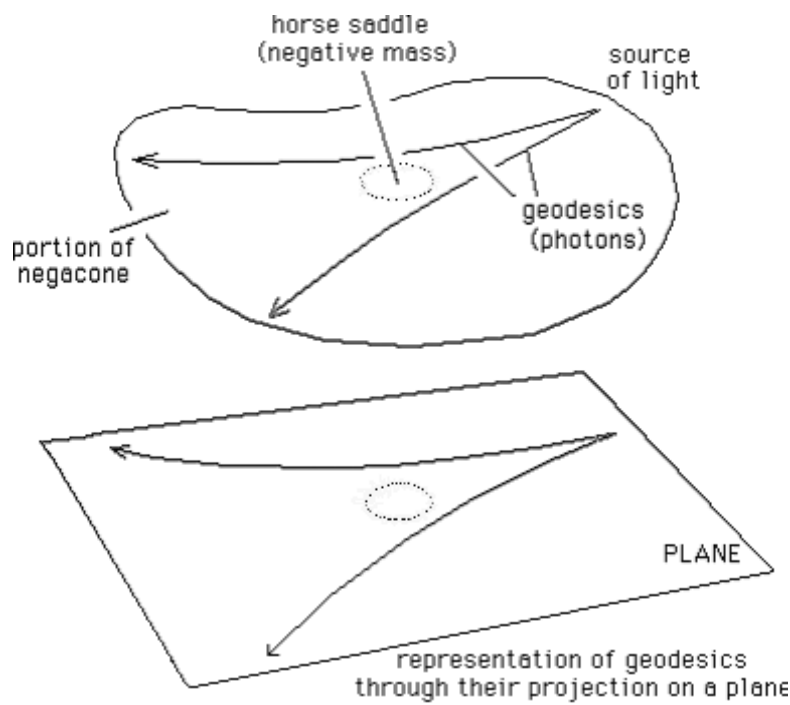

**Negative lensing due to an invisible negative mass concentration**

Theoretically, didactic images are not necessary for experienced geometers, but in practice even skilled geometers need mental picturing.

**References**

[1] J.M.Souriau, *Structure des Systèmes Dynamiques*, Editions Dunod, 1970. English translation in: *Structure of Dynamical Systems*, Birkhauser Ed. 1999
[2] J.M.Souriau : *Géométrie et relativité*. Hermann Ed. 1964
[3] A.Sakharov : "*CP violation and baryonic asymmetry of the Universe*". ZhETF Pis'ma **5** : 32-35 (1967) : Traduction JETP Lett. **5** : 24-27 (1967)
[4] A.Sakharov : "*A multisheet Cosmological Model*" Preprint Institute of Applied Mathematics, Moscow 1970
[5] A.Sakharov : "*Cosmological Model of the Universe with a time-vector inversion*". ZhETF **79** : 689-693 (1980) : Traduction in Sov. Phys. JETP **52** : 349-351 (1980)
[ 6 ] Linde. *Particle Physics and Inflationary Cosmology*, Harwood, Switzerland, 1990
[7] J.P.Petit : "*Univers énantiomorphes à flèches du temps opposés*", CRAS du 8 mai 1977, t.285 pp. 1217-1221
[8] J.P.Petit : "*Univers en interaction avec leur image dans le miroir du temps*". CRAS du 6 juin 1977, t. 284, série A, pp. 1413-1416
[9] J.P.Petit : Twin Universe Cosmology : Astronomy and Space Science 226 : 273-307, 1995
[10] J.P.Petit : The missing mass problem. Il Nuovo Cimento B Vol. 109 July 1994, pp. 697-710
[11] J.P. Petit, P.Midy & F.Landsheat : *Twin matter against dark matter*. Intern. Meet. on Atrophys. and Cosm. "Where is the matter?", Marseille 2001 june 25-29.
[12] J.P.Petit & G. D'Agostini : "*Bigravity. Bimetric model of the universe. 1 – An interpretation of the cosmic acceleration*". International Meeting on Variational Technics (CITV), Le Mont Dore, 2007 [http://arxiv.org/abs/0712.0067 ]